\documentclass[12pt]{article}
\usepackage{amsmath,amssymb,graphicx,mathrsfs,hyperref,slashed}
\newcommand{\be}{\begin{equation}}
\newcommand{\ee}{\end{equation}}
\newcommand{\bea}{\begin{eqnarray}}
\newcommand{\eea}{\end{eqnarray}}

\def\({\left(} \def\){\right)}

\renewcommand{\baselinestretch}{1.25}
\begin{document}
\title{\vspace{-1.8in}
{Quantum hair of black holes out of equilibrium}}
\author{\large Ram Brustein${}^{(1)}$,  A.J.M. Medved${}^{(2,3)}$
\\
\vspace{-.5in} \hspace{-1.5in} \vbox{
 \begin{flushleft}
  $^{\textrm{\normalsize
(1)\ Department of Physics, Ben-Gurion University,
    Beer-Sheva 84105, Israel}}$
$^{\textrm{\normalsize (2)\ Department of Physics \& Electronics, Rhodes University,
  Grahamstown 6140, South Africa}}$
$^{\textrm{\normalsize (3)\ National Institute for Theoretical Physics (NITheP), Western Cape 7602,
South Africa}}$
\\ \small \hspace{1.07in}
    ramyb@bgu.ac.il,\  j.medved@ru.ac.za
\end{flushleft}
}}
\date{}
\maketitle
\begin{abstract}
Classically, the  black hole (BH) horizon is completely opaque, hiding any clues about the state and very existence of its interior. Quantum mechanically and in equilibrium, the situation is not much different: Hawking radiation will now be emitted, but it comes out at an extremely slow rate, is thermal to a high degree of accuracy and thus carries a minimal amount of information about the quantum state within the BH. Here, it is shown that the situation is significantly different when a quantum BH is out of equilibrium. We argue that the BH can then emit ``supersized" Hawking radiation with a much larger amplitude than that emitted in equilibrium. The result is a new type of quantum hair that can reveal the state and composition of the BH interior to an external observer. Moreover, the frequency and amplitude of the new hair can be explained by the observer without invoking any new physical principles.  The new hair decays at a parametrically slow rate in comparison to the Schwarzschild time scale and can be detected through the emission of gravitational waves (and possibly other types of waves); for example, during and after a BH-merger event. The current discussion is motivated by a previous analysis, in the context of a recently proposed polymer model for the BH interior, that implies emissions just like those described here. We expect, however, that the new hair is a model-independent property of quantum BHs.
\end{abstract}
\newpage
\renewcommand{\baselinestretch}{1.5}\normalsize

\section{Introduction}

It was not too long ago when the classical picture of the
interior of a black hole (BH) was more or less accepted. However, thanks to the so-called firewall argument \cite{AMPS,MP} (and as recently reviewed in \cite{Mathur2017}) along with its less-celebrated forerunners \cite{Sunny,Mathur1,Braun},  a conflict between the classical description of a BH and the principles of quantum theory has been revealed.  As such, a variety of non-classical models of the interior have emerged as  leading candidates for a successor. (See, however, \cite{raju,pap}.)
But a consensus point of view on the correct description of the interior is still lacking, as evident from the  ongoing and rather intense debate (see \cite{harlow} for a summary). Some have suggested that the BH interior should, one way or another, be expelled from the accessible part of spacetime ({\em e.g.}, \cite{newhooft}). We, on the other hand, have suggested that the BH interior is composed of highly excited, interacting, long, closed strings
--- essentially, a ``ball of string'' or a collapsed polymer \cite{strungout}. Others have, however, proposed entirely different compositions for the interior
 ({\em e.g.}, \cite{nonpolymer}).

The recent detections of  gravitational waves (GWs)  from BH mergers \cite{LIGO} have elevated what was an  abstract academic debate about the laws of quantum gravity to a more tangible discussion about the expected signatures
in the GW data of either a non-empty or an excised interior \cite{Cardoso,noncardoso}. Meaning that each new proposal about the nature of BHs will have to confront such data  as it comes  in from subsequent merger observations.
Besides the resulting GWs, there could also be data from the emission of electromagnetic waves and neutrinos, although neither has been  detected so far.  Until now, the data has been completely consistent  with the predictions of classical general relativity \cite{pretorius,nonpretorius} (see, however, \cite{noncardoso,counterechoes}),
 but it is  too soon to reach any definitive conclusions.

We  have recently analyzed some of  the consequences of our proposed polymer description of the BH interior. In particular, it was argued  in \cite{collision} (also  see \cite{spinny}) that GW observations could  provide a means for distinguishing our model from that of a classical BH as well as from  other candidate models. The idea is that the interior matter of a polymer BH, which can be effectively  viewed as a fluid, will support pulsating modes in essentially the same way that a relativistic star does. These fluid modes would exist in addition to the standard spacetime modes of the exterior, and so their spectrum would then be added onto that  of the ring-down or quasinormal modes (QNMs) of a perturbed BH. The  polymer has an outer surface that behaves just like a BH horizon in the limit $\;\hbar\to 0\;$ but is otherwise only partially opaque \cite{emerge}. Models without such an effective horizon would likely have a spectrum that differs even more substantially from that  of a classical BH \cite{morecardoso}.

The bottom line is that a fluid-like description of the BH interior  gives rise to a new type of quantum hair, which  is emitted with  a parametrically lower frequency $\;\omega_I\sim v_I/R_S\;$ and a parametrically longer damping time  $\;\tau_I\sim R_S c/v_I^2\;$ \cite{collision} in comparison to the QNMs of its  classical counterpart. Here, $R_S$ is the Schwarzschild radius,  $c$ is the speed of light and $\;v_I < c\;$ is the velocity of sound for  a
fluid mode from  the $I^{\rm th}$ class.  For the polymer model in particular, the parametric difference is due to the introduction of a new scale, the string scale, and therefore a new dimensionless parameter $\;g_s=l_P/l_s\;$, the ratio of the Planck scale to the string scale. In this case, $\;v_I/c=g_s\;$ for
what would be the most experimentally accessible class of modes.

Here, we would like to discuss how a coupling between internal fluid modes and
emitted GWs (or other types of waves) can occur from the perspective of an observer on the outside. The external observer does have the prerogative of ignoring all knowledge about the interior but, then again,  should be able to  explain all phenomena in the framework of classical GR. Hawking has made this same point in \cite{info};
to wit, `All data on a ``hidden'' surface compatible with the observer's limited information are equally probable.' From this perspective, the interior of the BH is an imaginary construction whose sole justification is to serve as a mental crutch to help explain the properties of the emitted radiation. After all,  the BH horizon is supposed to prevent just such an emission as the interior is causally disconnected as far as this outside observer is concerned \cite{Mathur2017}.

Since the picture does seem sensible enough from an internal point of view \cite{collision,spinny}, what needs to be shown is that an external observer will attribute the source of the additional (fluid) modes to perturbations  of the exterior spacetime and not those  of the BH interior. Establishing this to be true is  the primary objective  of the current paper, and we are indeed able to confirm that the two perspectives are consistent.

The key to resolving the conflicting viewpoints is the realization that this external perspective for the fluid modes is really no more or less paradoxical than  that of Hawking radiation itself \cite{Hawk,info}. In spite of some arguments that the Hawking effect can be linked to mechanisms like pair production, quantum tunneling and so on, one can only learn about the interior indirectly by observations on the outside. And so, as the above  quotation correctly implies, any explanation of the Hawking process is just as viable as any other as long as its predictions are consistent with what is known or could be known about BHs.  Hence,  there has to be a vantage point for which the Hawking modes originate in the exterior spacetime because, just like for any other form  of matter, this radiation is not permitted to escape {\em through} the horizon.

What we will then argue is that, when viewed externally, the fluid modes are describing   ``supersized'' Hawking emissions. This is because each such event represents a large-amplitude coherent state of photons, gravitons, {\em etc.}
(akin to a electromagnetic or gravitational field) rather than a single boson. And, same as for the standard case, the supersized modes  must appear to have originated in the exterior spacetime. For either choice, regular or supersized, we will assert that this exterior picture is consistent  as long as the BH has, to some degree, deviated from its equilibrium state. The degree of deviation depends on the amount of energy that is injected into the specific mode and, therefore, also determines the amplitude of the respective  emission.

We will also  address  the puzzling absence of (damped) relativistic modes in the interior, which was a central finding in \cite{collision}. The question of interest is whether  this result  is an artifact of our particular model or  a physical consequence   of a more general nature.   We do find that this is indeed a general phenomenon, from both the internal and external perspectives.

Although relying on the results of a particular model, we expect that many of the  ideas and conclusions should apply just as well  to any ``BH-like'' object;
which is meant as an exotic spacetime containing exotic matter that can exist inside of an ultra-compact object while somehow resisting  gravitational collapse. This object should, simultaneously, exhibit all of the standard properties of a BH when viewed from the outside. Note that this excludes models lacking a ``horizon-like'' outer surface such as gravastars and wormholes.

The above claims are substantiated and further discussed in Section~2,
followed by a summary in Section~3.
Throughout the paper,  we ignore numerical factors of order one and fundamental constants are only made explicit when needed for clarity. For simplicity,  three large spacelike dimensions and
a non-rotating BH are assumed. When we refer to BHs (including the use
of the subscript $BH$),
the polymer model is implied unless stated otherwise.
 As the current results are often compared
to those in \cite{collision}, we follow this earlier treatment and
assume that the fluid modes are scalars.

\section{The external perspective}

An external observer  can, from her perspective, only see ``stuff'' which is
 on her side of the horizon. Whatever is supposed to be leaking out of the BH,
whether it be conventional or ``supersized" Hawking radiation,  must
have originated from outside the horizon as far as this observer is concerned. When the BH is close to its equilibrium state, this outside observer can use a ``horizon-locking gauge'' to describe the near-horizon geometry \cite{Poisson1,Poisson2}. In this gauge,  the equilibrium position of
the compact object's outer surface (or effective horizon)  stays at $\;r=R_S\;$ up to some high order in the relative strength of the perturbation.


\begin{figure}[htb]
\centerline{\includegraphics[width=5.5cm]{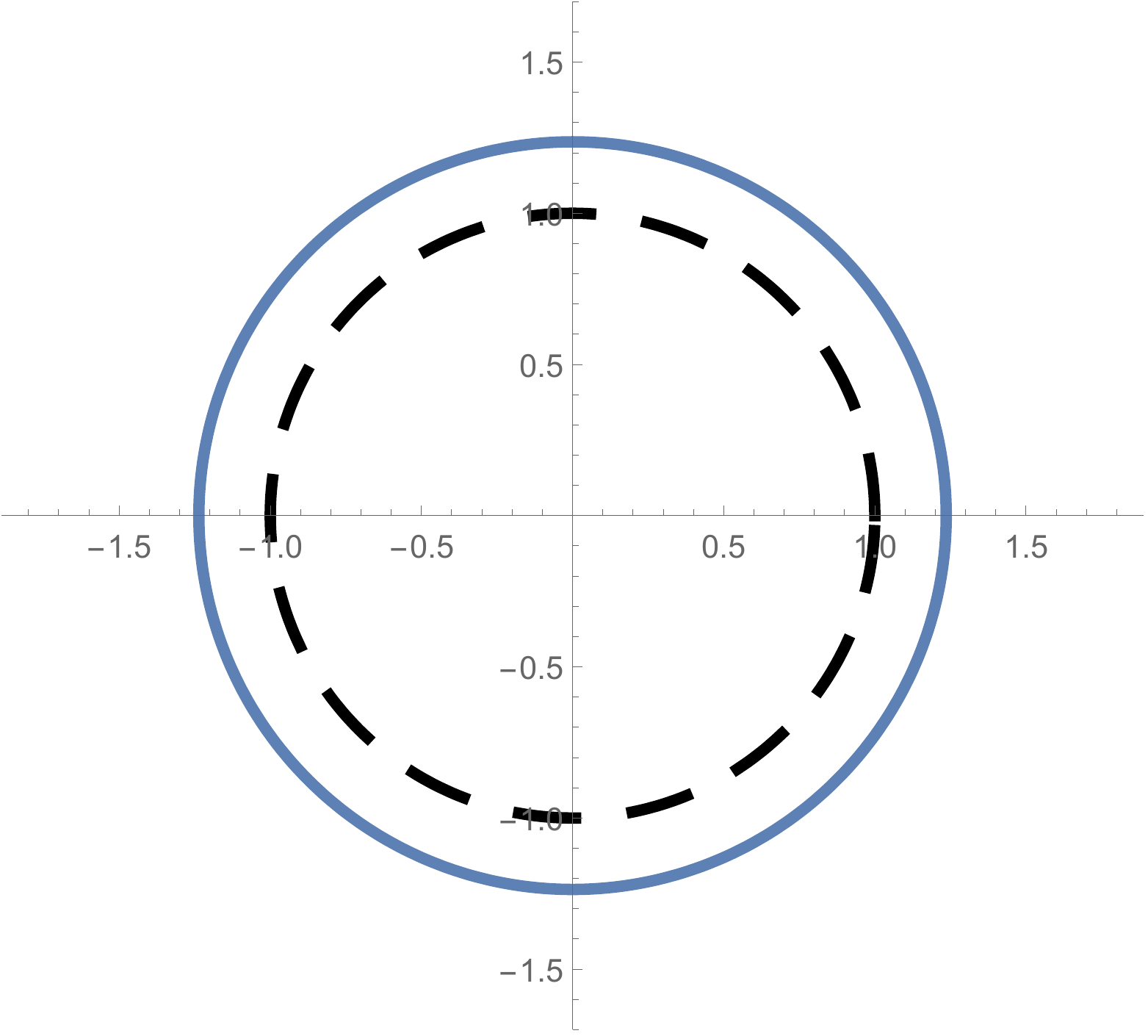}
\includegraphics[width=5.5cm]{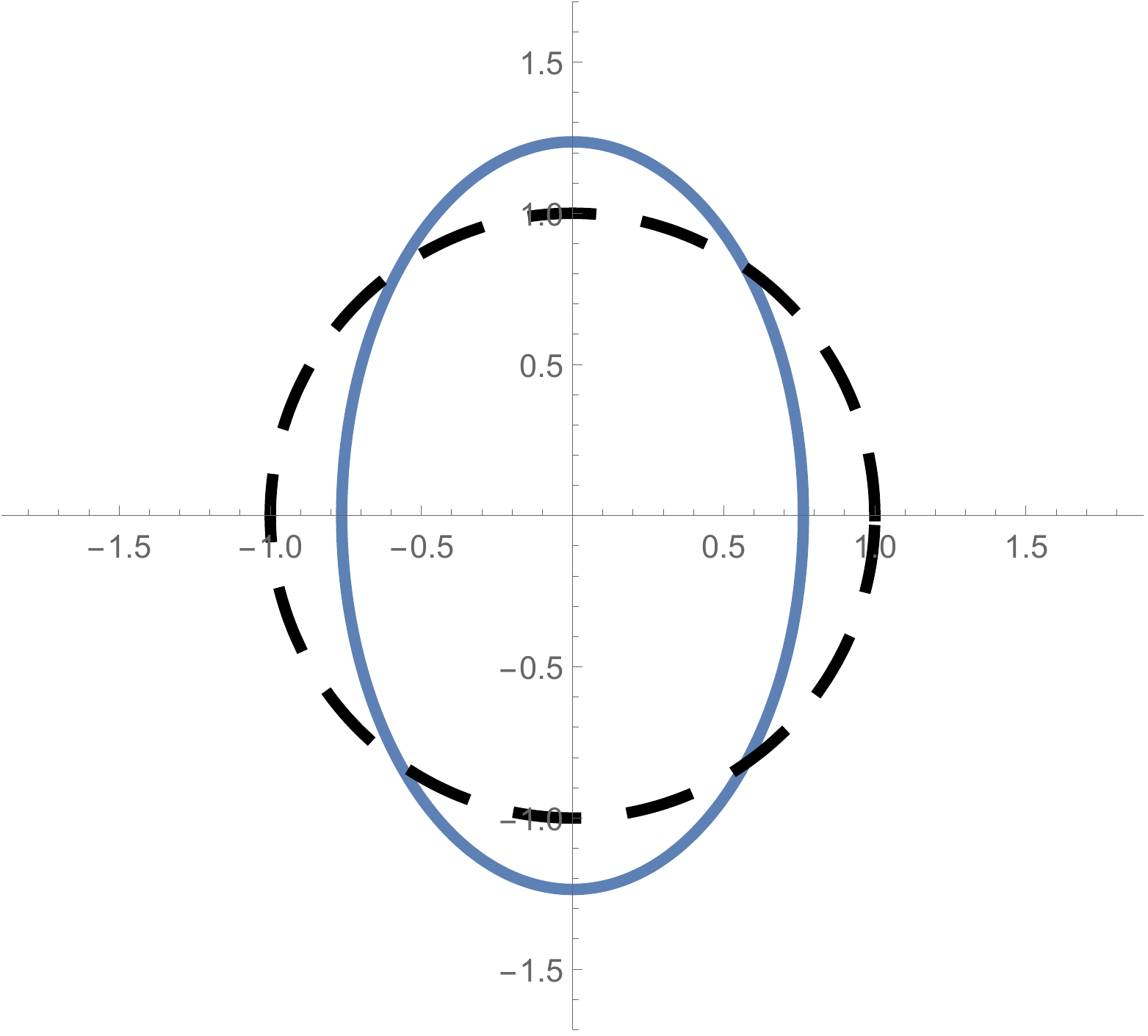}
\includegraphics[width=5.5cm]{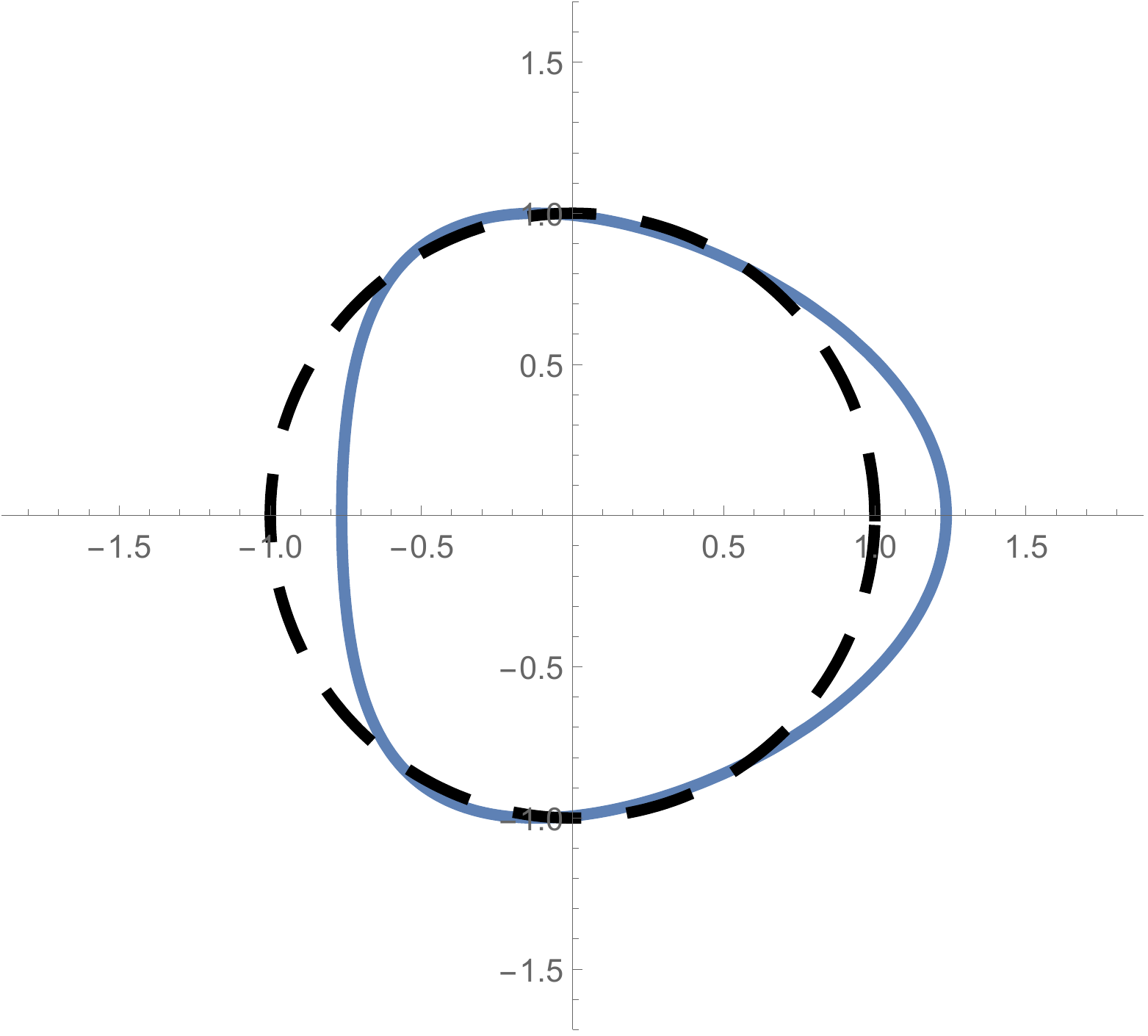}}
\caption{Visualization of the deformed horizon.  Scalar (left), dipole (center) and quadrupole (right) deformations are shown. The dashed, black circles depict the position of the unperturbed  horizon at $r=R_S$ and the solid, blue shapes depict its respective deformations.}
\label{fig:horizon}
\end{figure}

To understand how an outside observer would interpret the
supersized quantum radiation, it is necessary to know about  departures from equilibrium. To get a handle on this,  let us first recall a classical analogue: a tidal deformation of the horizon of a slowly rotating BH due to an external perturbation. This was first discussed by Hartle \cite{Hartle} and later by  O'Sullivan and Hughes \cite{OSH} (see, in particular, Appendix~B1 of \cite{OSH} and also \cite{Poisson1,Poisson2}),  who  visualized this setup by embedding a deformed sphere in a three-dimensional, flat Euclidean space.
The basic idea was to  lock up the position of  the sphere's outer surface
(as described above)
 and rather
interpret its deformation  as a perturbation in the
associated Ricci curvature. But one can just as well choose a gauge
for which the deformation is interpreted
as the  difference between the location of the outer surface and $R_S$.  This
difference  would, in our case, be  the extent that the internal fluid is either protruding out of or sinking into  the fiducial horizon.

Figure~1 can help one to  visualize the gravitational coupling between the deformed horizon and the external observer.  For example, a static quadrupole deformation of the horizon (rightmost panel, Fig.~1) changes the sphere from its unperturbed shape by an amount that scales with the strength of the perturbation times the second Legendre polynomial for the polar angle $\;P_2(\theta) \propto 3\cos^2{\theta}-1\;.$ More generally, the position and shape of the  deformed horizon can be expected to oscillate in time.

Perhaps contrary to expectations, the deformed surface can include both depressions and protrusions  irrespective of the direction of the perturbing force, as
illustrated in Fig.~1. Our interest is in places where the horizon is depressed inwards (equivalently, where the fluid protrudes outwards), as this implies that a portion of the interior has been momentarily ``exposed'' to the exterior spacetime. From an external point of view, such a depression of the horizon can only  be explained by the BH absorbing a flux of negative energy, just as the emission of standard Hawking radiation is  normally  explained \cite{BD}. However, the negative flux is really just a story that an outside observer has to   invent to  reconcile energy conservation with the flux of positive energy emanating out from the BH.

In the classical case of tidal horizon deformations, an outgoing flux occurs only for the superradiant modes of a  rotating BH. In the case of  Hawking radiation emerging from a polymer BH, the outgoing flux and compensating negative flux  are explained, internally, by a quantum effect that allows small loops of  string to break off  and detach from the string-filled interior. Supersized Hawking radiation should be similar but, in this case,  a large portion of string would  be detached collectively in a  short span of time.

Also from an internal perspective, the relative deformation $\Delta L/L$
 of the horizon
due to a particular restoring force (say the $I^{\rm th}$ one) is the ratio~\footnote{Unlike in \cite{collision}, we  now use the energy $E$ in place
of the free energy $F$, as these and their order-by-order corrections scale in parametrically the same
way in the polymer model.}
\be
\left.\frac{\Delta L}{L}\right|_I\; \sim\; \frac{(\Delta E)_I}{E}\;=\; \frac{p_I}{\rho}\;=\;
\frac{v^2_I}{c^2}\;,
\label{reldef}
\ee
where $v_{I}$ is again the sound velocity for the $I^{\rm th}$ mode, $p_I$ is its pressure  and $\;\rho\sim M_{BH}/R_S^3\;$ is the total energy density ($M_{BH}$
is the BH mass).  Here,  we have employed standard relations from thermodynamics, between stress and strain and between the pressure-to-energy-density  ratio and sound velocity.

Equation~(\ref{reldef}) can be used  to obtain an  expression for the redshift at the
outermost extent of the protruding fluid,
\be
\left.\sqrt{-g_{tt}}\right|_I\;=\; \sqrt{1-\frac{R_S}{r_I}} \;= \;\sqrt{1-\frac{R_S}{R_S\left(1+ \left.\frac{\Delta L}{L}\right|_I\right)}}\;\approx\;
\sqrt{\left.\frac{\Delta L}{L}\right|_I}
\;,
\label{commie1}
\ee
that is,
\be
\left.\sqrt{-g_{tt}}\right|_{I}\;\approx\; v_I/c\;.
\label{commie2}
\ee
The above estimates will be used  to determine the mode frequencies as  measured by an observer in the exterior, which will tell us if her observations
are consistent with those from an interior point of view.

For future reference, it should be noted that Eqs.~(\ref{commie1}) and (\ref{commie2}) are not compatible with a relativistic speed of sound $\;v_I=c\;$ since, in this case, $\Delta L/L$ cannot be small. This is the first indication that the limiting case of $\;v_I=c\;$ is problematic.

\subsection{Hawking radiation}

To illustrate the procedure of determining the external frequencies, it is useful  to start with the familiar case of standard Hawking radiation. Let us then begin with
\be
 (\Delta E)_{H}\;\sim\; T_{H}\;=\; \frac{1}{R_S}
\ee
and
\be
E\;=\;M_{BH}\;,
\ee
from which it  follows that
\be
\frac{(\Delta E)_{H}}{E}\;\sim\;\frac{1}{R_S M_{BH}}
\;=\; \frac{1}{S_{BH}}\;
\ee
and then
\be
\left.\sqrt{-g_{tt}}\right|_{H}\;=\;\sqrt{\frac{1}{S_{BH}}}\;=\;\frac{l_P}{R_S}\;,
\ee
where a subscript of $H$ indicates an associated property ({\em e.g.}, $T_H$ is the Hawking temperature), $S_{BH}$ is the BH entropy and $l_P$ is the Planck length.

This is the redshift at the location of the protruding fluid and, therefore, the location of the source as far as an external observer is concerned. But what frequency would this observer
assign to a Hawking mode at the same point?
It is natural to ascribe a wavelength of $l_P$ to a  near-horizon
Hawking mode. The logic here follows that of  't Hooft's ``brick-wall'' model \cite{BW}. Formally, one can start with
$\;(\Delta L)_H = \frac{L}{E} (\Delta E)_H= \frac{R_S}{S_{BH}}=\frac{l^2_P}{R_S}\;$,
which corresponds to a proper  length of $l_P$.
And so the observer assigns this mode  with a frequency at the
source of $\;\omega^{(H)}_{source}=c/l_P\;$.
The frequency at the location of the external observer
is then found  by redshifting  its value at the source. This process
yields the expected result,
\be
\omega^{(H)}_{ext}\;= \;\omega^{(H)}_{source}\left.\sqrt{-g_{tt}}\right|_{H}
\;=\;
T_{H}\;.
\ee

This is not exactly groundbreaking physics, as the above argument runs along the same lines as that of the membrane paradigm \cite{TPM}. The difference here, though, is that we did not have to conjecture a location for the (so-called) stretched horizon before determining the redshift.

The polymer model realizes the same value for the Hawking temperature from an internal perspective  \cite{emerge}. A small loop of string which has broken off from one of the typically long loops in what is a bound state of interacting, highly excited, closed strings  will have some probability of escape, and a calculation reveals that both the rate and energy of emission agree with $T_H$.  In this way, consistency between the exterior and interior perspectives has been established, at least as far as it concerns the case of conventional Hawking radiation.

\subsection{``Supersized" Hawking radiation}

\subsubsection{Frequency of emission}

Let us next consider some non-relativistic fluid mode, beginning  with
its frequency as seen by an external observer.
We already know that the redshift  at the location of the  protruding fluid
(which is exterior to  but still in the vicinity of $\;r=R_S$)
is $v_I/c$, and so  it becomes prudent to ask
 about the mode frequency at this same  ``source'' location. An external observer, who is unaware of the fluid, would confuse  these non-relativistic fluid
modes with relativistic spacetime QNMs, for which the
wavelength at the  source
 would be  approximately $R_S$. She would therefore assign them a frequency at the source of
\be
\omega^{(I)}_{source}\;=\;\frac{c}{R_S}\;,
\ee
from which one can  deduce that
\be
\omega^{(I)}_{ext}\;= \;\omega^{(I)}_{source}\left.\sqrt{-g_{tt}}\right|_{I}
\;=\; \frac{v_I}{R_S}\;,
\ee
again as expected (see the Introduction). The difference between the internal and external perspectives can then only be one of interpretation.

Since the frequencies redshift, one might wonder why the energies $(\Delta E)_I$ in Eq.~(\ref{reldef})
do not. In reality they do but, in both cases (standard and supersized),
the values that were used for $(\Delta E)_I$ and $E$ are already what would be  measured by an asymptotic observer. We know this  because the standard Hawking case
can be used to calibrate all other cases.   The two energies $(\Delta E)_I$ and $E$ at the source would then be blueshifted from their asymptotic values  in  the same way, leaving their ratio undisturbed.

\subsubsection{Coupling to the exterior and decay time}

To complete our consistency check, the estimates from \cite{collision} for an emitted  energy
of  $\;(\Delta E)_I = v_I^2 E_I\;$ and a damping time of  $\;\tau_I= R_S/v_I^2\;$
need to be
similarly reproduced from an external perspective. Here, $E_I$ is the amount of energy which has been injected into the $I^{\rm th}$ mode by
the deforming force. (It was assumed in \cite{collision}
that $\;E_I\sim M_{BH}$.)

For an exterior observer, the supersized Hawking radiation is relativistic and has a frequency of  $\;\omega_I=v_I/R_S\;$. Therefore, she must conclude that the wavelength of the radiation at distances far away from the horizon is $\lambda_I=R_S c/v_I$. The same conclusion can be arrived at by the fact that, like  before, the wavelength near  the source  must be $\lambda\sim R_S$, which then
asymptotically redshifts to $\;\lambda_I\sim R_S c/v_I\;$.

Now consider that the same observer  attributes the source with  a radial size
of  about $R_S$. She then just needs to know that the transmission cross-section for such long wavelength modes through a proportionally smaller surface of area $A$ is determined by the ratio $A/\lambda^2$, which translates into $\;R_S^2/\lambda_I^2= v_I^2\;$ for the case at hand. We can  conclude that the coupling or efficiency of emission goes as $v_I^2$, so that the energy in the emitted wave scales as
\be
(\Delta E)_I\;\sim\; E_{I} v_I^2\;,
\ee
in agreement with the internal perspective \cite{collision}. This is based on the assumption that most of the mode energy is being emitted in the form of coherent waves rather than  dissipating as heat.

Meanwhile, the damping time for any given mode $\tau_I$ is directly related to the
corresponding relaxation time of the BH. The latter can be deduced with
an inspection of
\be
\frac{dE_I}{dt}\;\propto\; (\Delta E)_I \;\sim\; v_I^2 E_I\;,
\label{taudamp}
\ee
which implies a relaxation time that scales with the inverse of  $v_I^2$
and likewise for the damping time $\tau_I$.
One is then led to the expected result of
$\;\tau_I \sim R_S/v_I^2\;$, where the factor of $R_S$ follows simply for dimensional  reasons and the knowledge that the Schwarzschild time is the only classically available time scale.

In summary, the supersized Hawking radiation oscillates with a frequency of $\;\omega_{I}=v_I/R_S\;$, carries away an energy of $\;(\Delta E)_I = E_I v_I^2\;$
(where it is expected that  $\;E_I\sim M_{BH}$)
 and decays with a characteristic time of $\tau_{I}=R_S/v_I^2\;$. This can be compared to
the standard Hawking emissions  with a frequency of $\;\omega_H=1/R_S\;$, an effective coupling
of \;$v^2_H=1\;$ ({\em cf},  $\;\omega_{I}=v_I/R_S$)  and an emitted energy of $1/R_S$, leading to a decay time of $\;\tau_H=R_S/v_H^2=R_S\;$, again just as expected.

\subsubsection{Coupling to gravitational waves}

Also of interest is the  strength of the coupling of the fluid modes to  external GWs, as
this along with $(\Delta E)_I$ is what determines the amplitude of the emitted GWs. The coupling strength can be determined using Einstein's quadrupole formula
\be
\langle h \rangle \;\sim\; \frac{1}{r} \ddot{Q}\;.
\ee
This means that, for an external observer,
\be
\langle h \rangle_I
\;\sim\;
\ddot{Q}_I
\;\sim\;
(\Delta E)_I R_S ^2\omega_I^2\;\sim\;
(E_I v_I^2) (R_S\omega_I)^2
\;\sim\;
 E_I v_I^4\;,
\label{hamp}
\ee
where the factor of  $R_S^2$  can be  attributed to the quadrupole moment of the emitting object and a dot denotes a time-derivative. Also, the two factors of frequency are due to the pair of time derivatives,  which further suppresses  the amplitude of the emitted GWs. Equation~(\ref{hamp}) agrees with the internal version of the same calculation \cite{collision}.

\subsubsection{Absence of relativistic modes}

From an internal perspective, the absence of  relativistic fluid modes can be traced to  the  polymer being near its equilibrium state and  an incompatibility between the two boundary conditions that any fluid mode is required to satisfy:
vanishing at the center of the object and outgoing at its surface.
Moreover, the leading correction to the  (free)  energy has to be parametrically small, $\;\Delta E/E< 1\;$ (see Eq.~(\ref{reldef}) and  the comment  just before Subsection~2.1), from which it  follows that $\;v_I^2/c^2 = \Delta E/E < 1\;$.

From an external perspective, it is rather the continuity of the emission  at $\;v_I=1\;$ which  makes the emission of such waves impossible. This is because  $\;v_I>1\;$ is unphysical and therefore unacceptable,  the amplitude of
such faster-than-light waves  has to  vanish identically. The condition of continuity then implies that the amplitude of waves for which  $\;v_I=1\;$ must similarly vanish.

Let us explain the continuity argument in a more detailed way: An external relativistic mode could never have been redshifted, as this would imply that it had been sourced  by a fluid mode whose  sound velocity was faster than the  speed of light. Now consider that, for a (would-be) relativistic fluid  mode,  $\;\omega= \alpha c/R_S\;$ and $\;\lambda = R_S/\alpha\;$, where  $\alpha$ is some constant of order 1 which takes into account any  neglected numerical factors.  But if its wavelength is indeed $\;R_S/\alpha\simeq R_S\;$, this mode  must have originated somewhere close to the horizon and must then  have experienced  a significant redshift. Conversely, to suffer no redshift, it would have to be produced far away from the horizon with  a wavelength that is parametrically larger than $R_S$. Such a  mode could not possibly be under the influence of the BH and so ---  even if it somehow defied the condition of continuity and  did exist ---  an external observer would not consider it to be  part of the BH's QNM spectrum. This argument does not preclude the existence of the standard class of relativistic spacetime  QNMs, as these are a consequence of  waves in the exterior spacetime and not of fluid modes from inside the BH.

\subsubsection{The potential barrier}

Finally, we would now like to show that the gravitational potential barrier at about $\frac{3}{2}R_S$ does not affect in any significant way the emission of the supersized Hawking radiation; an  assumption that  was implicit
in \cite{collision}.

To understand this claim, let us consider  a massless particle with a modest angular momentum; then the peak in the barrier goes as $1/R_S$ when expressed in units of energy (rather than units of energy squared as it normally appears). That the peak is of the same order as $T_{H}$ is what explains the famous grey-body factors affecting the emission of the standard Hawking radiation ({\em e.g.}, \cite{harlow}).
On the other hand, the energy of a supersized emission is of order $\;M_{BH} v_I^2$ for scalar modes and $\;M_{BH} v_I^4$ for gravitons, as $\;E_I\sim M_{BH}\;$
can be expected. Meaning that the ratio of the radiated energy to the height of
the barrier is  $\;M_{BH} v_I^4/(1/R_S)= S_{BH} v_I^4 \gg 1\;$ for the
fluid modes of interest ({\em i.e.}, those for which the resulting
GWs could be experimentally detected). A supersized emission, which is really a large coherent state of gravitons, will not be affected by barrier at all. This has become  a classical problem in which the energy of the wave far exceeds that of the potential barrier.

\section{Conclusion}

It was shown that, if an ultra-compact object is non-empty but does have a surface that acts effectively as a BH horizon,  interior modes can nevertheless couple to emitted GWs or, for that matter, other types of waves (such  as  electromagnetic waves and neutrinos). An external observer will view the interior modes as supersized Hawking emissions which originated close to but {\em outside} the equilibrium position of the  horizon. Moreover, we have shown that the same point of view applies just as well to standard Hawking radiation.

Although these conclusions rely on the intuition gained from studying the polymer model for the BH interior \cite{collision} (also \cite{spinny}), we believe that they are not specific to the polymer model and would readily carry over to any ultra-compact object containing non-trivial fluid-like matter and having an outer surface that acts like a BH horizon to some level of approximation.

The resulting picture is suggestive of a new type of  BH hair for which a parametrically
long time of shedding is required. In fact, the existence of novel  BH hair should be  part and parcel for any BH-like object  containing non-trivial matter and could yet be the key which unlocks  the door to
the secretive world behind the horizon. Such revelations could come about
through the observation of  GWs resulting from BH mergers, hopefully in the near future.

We have also  addressed  the  absence  of  damped relativistic modes in the interior, even though these are  ubiquitous in  some of the analogous calculations for relativistic stars. There is, however, some evidence
that the suppression of  relativistic fluid modes is a  more general phenomenon
({\em e.g.}, \cite{inversecowling,Kokk1}). If so, our analysis could prove
helpful in a broader range of studies.

\section*{Acknowledgments}

The research of RB was supported by the Israel Science Foundation grant no. 1294/16. The research of AJMM received support from an NRF Incentive Funding Grant 85353 and an NRF Competitive Programme Grant 93595. AJMM thanks Ben Gurion University for their  hospitality during his visit.

\end{document}